\documentclass[preprint,showpacs,preprintnumbers,amsmath,amssymb]{revtex4}

\usepackage{etex}
\usepackage{graphicx}% Include figure files
\usepackage{dcolumn}% Align table columns on decimal point
\usepackage{bm}% bold math
\usepackage{amssymb,amsthm,amscd,amsbsy,array}
\usepackage{amsmath,dsfont}
\usepackage{soul} 
\usepackage{graphics,xcolor}

\numberwithin{equation}{section}

\newcommand{\half}{{\frac{1}{2}}}
\def\2{{\half}}
\newcommand{\const}{\mathop{\rm const}\nolimits}
\def\bA{{\bm{A}}}

\def\bR{{\mathds{R}}}
\def\bbeta{{\bm{\beta}}}

\def\bgamma{{\bm{\gamma}}}

\def\bomega{{\bm{\omega}}}
\def\bSigma{{\bm{\Sigma}}}
\def\bp{{\bm{p}}}

\def\tbP{\widetilde{\bm{P}}}
\def\tbx{\tilde{\bm{x}}}
\def\tbX{\widetilde{\bm{X}}}

\def\ba{{\bm{a}}}

\def\bnabla{\mbox{\boldmath$\nabla$}}
\def\br{{\bm{r}}}

\def\bP{{\bm{P}}}

\def\bB{{\bm{B}}}
\def\bb{{\bm{b}}}

\def\bnabla{{\bm{\nabla}}}

\def\bp{{\bm{p}}}
\def\bP{{\bm{P}}}

\def\bs{{\bm{s}}}

\def\hbp{{\widehat{\bm{p}}}}
\def\hbb{{\widehat{\bm{b}}}}
\def\bx{{\bm{x}}}
\def\bX{{\bm{X}}}

\def\bW{{\bm{W}}}

\def\bX{{\bm{X}}}

\def\beq{\begin{equation}}
\def\eeq{\end{equation}}
\def\beqa{\begin{eqnarray}}
\def\eeqa{\end{eqnarray}}

\def\barray{\left(\begin{array}}
\def\earray{\end{array}\right)}
\def\barraynb{\begin{array}}
\def\earraynb{\end{array}}
\def\IR{{\mathds{R}}} %%%%% Reals

\def\so{{\rm so}}
\def\SO{{\rm SO}}
\def\ort{{\mathfrak{o}}}

\def\smallover#1/#2{\hbox{$\textstyle\frac{#1}{#2}$}} %

\newcommand{\bbR}{\mathbb{R}}

\newcommand{\fg}{{\mathfrak{{g}}}}

\newcommand{\be}{{\mathbf{e}}}

\newcommand{\Coad}{{\mathrm{Coad}}}

\usepackage{color}

\begin{document}

\preprint{arXiv~:~1411.6541v3
%\red{WS-v3-7P}
}

\title{Wigner-Souriau translations \\and\\ Lorentz symmetry of chiral fermions
}

\author{
C. Duval$^{1}$\footnote{mailto:duval@cpt.univ-mrs.fr},
M. Elbistan$^{2}$\footnote{mailto:elbistan@itu.edu.tr},
P. A. Horv\'athy$^{2,3}$\footnote{mailto:horvathy@lmpt.univ-tours.fr},
P.-M. Zhang$^{3}$\footnote{e-mail:zhpm@impcas.ac.cn}
}

\affiliation{
Aix-Marseille University, CNRS UMR-7332,  Univ. Sud Toulon-Var
13288 Marseille Cedex 9,
(France)
%Canonical affiliation below:
%Aix-Marseille Universit\'e, CNRS, CPT, UMR 7332, 13288 Marseille, France.\\
%Universit\'e de Toulon, CNRS, CPT, UMR 7332, 83957 La Garde, France.
\\
${}^2$Laboratoire de Math\'ematiques et de Physique
Th\'eorique,
Universit\'e de Tours,  
(France)
\\
$^3$Institute of Modern Physics, Chinese Academy of Sciences, Lanzhou, (China) 
}

\date{\today}

\begin{abstract}
Chiral fermions can be embedded into Souriau's massless spinning  particle model by ``en\-slaving'' the spin, viewed as a gauge constraint. The latter is not invariant  under Lorentz boosts; spin enslave\-ment can be restored, however, by a Wigner-Souriau (WS) translation, analo\-gous to a compensating gauge transformation. The combined transformation is precisely  the  recent\-ly  uncovered twisted boost, which we now extend to finite transformations. WS-translations are identified with the stability group of a motion acting on the right on the Poincar\'e group, whereas the natural Poincar\'e action corresponds to action on the left. The relation to non-commutative mechanics is explained.
%\red{\bf WS-v3-7P}
\end{abstract}


\maketitle

%\tableofcontents

%%%%%%%%%%%%%%%%%%%%%%%%%%%%%%%%%%%%%%%%%%%%%%%%%%%%%%%%%%%%%%%%%%%%%%%%%%%%%%
%%%%%%%%%%%%%%%%%%%%%%%%%%%%%%%%%%%%%%%%%%%%%%%%%%%%%%%%%%%%%%%%%%%%%%%%%%%%%%
\section{Introduction}
%%%%%%%%%%%%%%%%%%%%%%%%%%%%%%%%%%%%%%%%%%%%%%%%%%%%%%%%%%%%%%%%%%%%%%%%%%%%%%
%%%%%%%%%%%%%%%%%%%%%%%%%%%%%%%%%%%%%%%%%%%%%%%%%%%%%%%%%%%%%%%%%%%%%%%%%%%%%%

Semiclassical chiral fermions of spin $\half$ can be described by the phase-space action 
\beq
S=\int\Big(\big(\bp+e\bA\big)\cdot\frac{d{\bx}}{dt}-
h
-\ba\cdot\frac{d{\bp}}{dt}
\Big)dt,
\qquad 
h=|\bp|+e\phi(\bx),
\label{chiract}
\eeq
where
 $\ba(\bp)$ is a vector potential for 
the ``Berry monopole'' in $\bp$-space,
$
\bnabla_{\bp}\times\ba=\displaystyle\frac{\hbp}{2|\bp|^2},
$
$\hbp=\displaystyle{\bp}/{|\bp|}$, $\bp\neq0$. 
Here  $\bA(\bx)$ and $\phi(\bx)$ are [static] vector and scalar potentials and $e$ is the electric charge \cite{Stephanov,SonYama2,Stone,Stone3,ChenSon,DHchiral}.
A distinctive feature  is that spin is ``enslaved'' to the momentum, i.e., 
\beq
\bs= \half\, \hbp.
\label{enslavement}
\eeq
An intriguing aspect of the model is its \emph{lack of manifest Lorentz symmetry}. Recently \cite{ChenSon},  it was shown, though, that modifying the dispersion relation in (\ref{chiract})
as $\displaystyle{}h=|\bp|+e\phi(\bx)+\frac{\hbp\cdot\bB}{2|\bp|}$
yields a theory which is covariant w.r.t.
Lorentz transformations. Turning off the external field, their expression \# (6) reduces to
\begin{eqnarray}
\label{twistedboost}
\delta\bx=\bbeta\,t+
\displaystyle\bbeta\times\frac{\hbp}{2|\bp|},
\qquad
\delta\bp=|\bp|\,\bbeta,
\qquad
\delta{t}=\bbeta\cdot\bx,
\end{eqnarray}
where $\bbeta$ is an infinitesimal Lorentz boost.
This formula has  also been found, independently
\cite{DHchiral}, by relating the chiral fermion to Souriau's model of a relativistic massless spinning particle \cite{SSD,DHchiral}. 

The chiral and the Souriau systems \cite{SSD} have seemingly different degrees of freedom: the first one has ``enslaved'' spin, while the latter has   ``unchained'' spin represented by a vector $\bs$, which may not satisfy (\ref{enslavement}). However, the  freedom specific to the massless case of applying what we call a \emph{Wigner-Souriau} (WS) \emph{translation}, Eq. (\ref{finiteWigner}) below  \footnote{Wigner-Souriau translations (called ``$Z$-shifts'' in \cite{DHchiral}) were known to Wigner \cite{Wigner} and to Trautmann \cite{Trautman}, Penrose \cite{Penrose}. Their use in the semiclassical framework was advocated by Souriau \cite{SSD}. They were overlooked by most present authors with the notable exception of Stone et al. \cite{Stone3}, who also suggested viewing them as gauge transformations.},
allows one to eliminate the additional degrees of freedom of the Souriau framework, so that the two systems have identical spaces of motions \cite{SSD,DHchiral}.
This enabled us to \emph{``export''} the natural Lorentz symmetry to the chiral system, yielding (\ref{twistedboost}).

In this Note we derive (\ref{twistedboost}) in another, rather more intuitive way, namely by \emph{embedding} the chiral fermion model directly into the Souriau model, namely by viewing spin enslavement,  (\ref{enslavement}),
as a \emph{gauge condition} with WS translations 
and viewed as  gauge transformations. 
Then 
our clue is that a natural Lorentz boost does not leave the constraint~(\ref{enslavement}) invariant; the ``gauge'' condition (\ref{enslavement}) can, however, be restored by a WS translation, yielding,  once again, (\ref{twistedboost}).
This is reminiscent of what happens in \emph{gauge theories}, where 
a space-time transformation can  be a symmetry when the variation of the vector potential can be compensated by a suitable gauge transformation \cite{FoMa}. 

Further insight is gained in
Section \ref{Wigner} which clarifies the geometry hidden behind~:
while the natural Poincar\'e action corresponds the left-action of the Poincar\'e group on itself, WS translations correspond to the right-action of the stability group of a motion.

We conclude our Letter by some remarks about non-commutativity.

%%%%%%%%%%%%%%%%%%%%%%%%%%%%%%%%%%%%%%%%%%%%%%%%%%%%%%%%%%%%%%%%%%%%%%%%%%%%%%
\section{Symplectic description of the chiral and  the massless spinning models}\label{symplecticgSec}
%%%%%%%%%%%%%%%%%%%%%%%%%%%%%%%%%%%%%%%%%%%%%%%%%%%%%%%%%%%%%%%%%%%%%%%%%%%%%%

Both the chiral and the Souriau models can  conveniently be described within Souriau's framework \cite{SSD,DHchiral}, where the classical motions are identified with curves or surfaces in some evolution space, and are tangent to the kernel of a closed two-form; see \cite{DHchiral} for details. 
We limit our considerations to the free case; coupling to external fields has been  discussed in the literature, see \cite{DHchiral,ZHanom}. 

%%%%%%%%%%%%%%%%%%%%%%%%%%%
%\subsection{Chiral motion}
%%%%%%%%%%%%%%%%%%%%%%%%%%%

We first consider the [free] chiral model (\ref{chiract}). It has been shown \cite{DHchiral} that the associated variational problem admits an alternative geometric formulation.
To that end, we introduce the seven dimensional {evo\-lution space} $V^7=T(\IR^3\!\setminus\!\{0\})\times \IR$ described by  triples $(\bx,\bp,t)$, and endow it with the two-form $\sigma$ defined by
\beqa
\sigma=\omega-dh\wedge dt
\quad\hbox{where}\quad 
\omega=dp_i\wedge dx^i-\frac{1}{4|\bp|^3}{}
\epsilon^{ijk}\,p_i\,dp_j\wedge dp_k\,,
\quad
h=|\bp|.
\quad
\label{freechirsymp}
\eeqa
The two-forms $\omega$ and $\sigma$ are closed, since $\bnabla_{\bp}\cdot\bb=0$.
The  kernel of $\sigma$ defines an
integrable distribution, whose leaves [integral manifolds] can be viewed as
generalized solutions of the variational problem. Here, the kernel is one-dimensional and a curve 
$(\bx(t),\bp(t),t\big)$ is tangent to it  iff
 the equations of motion,
\beq
\displaystyle\frac{d{\bx}}{dt}=\hbp,
\qquad
\displaystyle\frac{d{\bp}}{dt}=0,
\label{chireqmot}
\eeq
 are satisfied \cite{DHchiral}; the solution is plainly
$ 
\bp=\bp_0=\const,
\,
\bx(t)=\bx_0+\hbp\,t,
 \,
\bx_0=\const,
$ 
i.e., the motion is in the $\hbp$ direction with the velocity of light.

%%%%%%%%%%%%%%%%%%%%%%%%%%%%%%%%%%%%%%%%%%%%%%%%%%%%%%
%\subsection{Motion of a massless spinning particle}
%%%%%%%%%%%%%%%%%%%%%%%%%%%%%%%%%%%%%%%%%%%%%%%%%%%%%%

The Souriau model admits a similar description \cite{DHchiral}. Restricting ourselves again to the free case, the
evolution space  here is $9$-dimensional and is described by 
\beq
V^9=\left\{
R,P\in{\bbR}^{3,1}, S\in\ort(3,1)\,\strut\Big\vert\,P_\mu{}P^\mu=0,P^4>0,S_{\mu\nu}P^\nu=0,{S}_{\mu\nu}{S}^{\mu\nu}=\half\right\}
\label{spinevspace}
\eeq
The evolution space
is endowed with the closed two-form 
\begin{equation}
\sigma
=
-dP_\mu\wedge{}dR^\mu
-2\,d{S}^\mu_{\;\lambda}\wedge{S}^\lambda_{\;\rho}\,d{S}^\rho_{\;\mu}.
\label{freespinsigma}
\end{equation}
A $(3+1)$-decomposition can be introduced by writing, in a Lorentz frame, $R=(\br,t)$, and $P=(\bp,|\bp|)$ where $\bp\neq0$. 
The components ${S}_{\mu\nu}$ of the spin tensor can in turn be split into space and space-time components, 
\beqa
S_{ij}=\epsilon_{ijk}\,s^k,
\quad
S_{j4}=\Sigma_j
\quad\hbox{where}\quad
\bSigma=\hbp\times\bs,
\label{Spq}
\eeqa
the $3$-vector $\bs$ being interpreted as the spin in the chosen Lorentz frame.
Note that $\bSigma$ is not an independent variable, so that 
a point of $V^9$ can  be labeled by
$(\br,t,\bp,\bs)$. 
Then the [free] equations of motion  
associated with the kernel of $\sigma$ in (\ref{freespinsigma}) are expressed as \footnote{These equations can also be obtained in a Wess-Zumino framework \cite{Skagerstam}, namely as the variational equations of the Lagrangian \# (3.1) of \cite{Skagerstam}; the latter is in fact the one-form (B.1) in~\cite{DHchiral}, whose exterior derivative is the two-form $d\big(\mu_0\cdot{}g^{-1}dg\big)$. % in (\ref{G2form}).
}, 
\beq
-\bp\cdot\dot{\br}+|\bp|\,\dot{t}
=0,
\qquad
\dot{\bp}=0,
\qquad
\dot{\bs}=\bp\times\dot{\br},
\label{3+1eqmot}
\eeq
and can be deduced from Eq. (3.6) in \cite{DHchiral}.
A particular solution of (\ref{3+1eqmot}) is obtained by \emph{embedding the chiral solution 
 above into the spin-extended evolution space $V^9$} by identifying $\br$  with  $\bx$ in (\ref{chireqmot})
and  completing [trivially] with a constant spin vector,
$ 
\bp=\bp_0=\const,
\,
\br(t)=\br_0+\hbp\,t,
\,
\br_0=\const.,
\,
\bs(t)=\bs_0=\const.
$ 
A remarkable feature of Eqs (\ref{3+1eqmot}) is that, for an arbitrary $3$-vector $\bW$, the transformation we 
refer to as a \emph{Wigner-Souriau} (WS) \emph{translation}  \cite{Wigner, Trautman, Penrose, SSD, Stone3, DHchiral},
\beq   
\br\to\br+\bW,
\,
t\to t+\hbp\cdot\bW,
\,
\bp\to\bp,
\,
\bs\to\bs+\bp\times\bW
\label{finiteWigner}
\eeq 
takes a solution of (\ref{3+1eqmot}) into another, equivalent one. 
 The kernel of (\ref{freespinsigma}) is invariant under WS-translations and is in fact $3$-dimensional, swept by the images of the embedded solutions. 

The spin vector here, $\bs$, is not necessarily ``enslaved'', i.e., may not be parallel to the momentum, $\bp$.
The spin constraint in (\ref{spinevspace}) implies nevertheless that the projection of the spin onto the momentum and the perpendicular component $\bSigma$
satisfy, 
\beq
\bs\cdot\hbp
=\half\,,
\qquad
\bSigma=\hbp\times\bs,
\label{projconst}
\eeq
respectively, see (\ref{Spq}). 
From the WS action above we infer that 
$ 
\bSigma \to \bSigma +\hbp\times(\bp\times\bW)
$. It follows that
 $\bSigma$ can be \emph{eliminated}~: choosing
 $\bW=\bSigma/|\bp|$ carries $\bSigma$ to zero.

The Poincar\'e group acts naturally on $V^9$,
namely according to \cite{SSD,DHchiral}
\beq
\left\{\barraynb{lll}
 \delta\br&=&\bomega\times\br+\bbeta{}t+\bgamma,
\\
\delta t&=&\bbeta\cdot\br+\varepsilon,
\\
\delta\bp&=&\bomega\times\bp+\bbeta\,|\bp|,
\\
\delta\bs&=& \bomega\times\bs-\bbeta\times\bSigma,
\earraynb\right.
\label{infPVrtps}
\eeq 
where $\omega,\bbeta,\bgamma,\varepsilon$ are identified with infinitesimal rotations, boosts, translations and time-translations; their action duly projects to Minkowski space-time as the natural one. In what follows, we focus our attention at boosts;
WS-translations will be studied further in Sec. \ref{Wigner}.

%%%%%%%%%%%%%%%%%%%%%%%%%%%%%%%%%%%%%%%%%%%%%%%%%%%%%%
\section{Embedding the chiral system into the massless spinning model}\label{embeddingSec}
%%%%%%%%%%%%%%%%%%%%%%%%%%%%%%%%%%%%%%%%%%%%%%%%%%%%%%%%%%%%%

Now we embed the evolution space of the chiral model,  $V^7$, into that of the massless spinning particle, $V^9$. We note first that, by (\ref{projconst}),  spin enslavement,
(\ref{enslavement}), is  equivalent to 
\beq
\bSigma=
 \hbp\times\bs=0,
\label{enslavement2}
\eeq
which, viewed as a constraint, 
defines a seven dimensional submanifold  of $V^9$ that we 
parametrize with $\br,\bp,t$  and  denote (with a suggestive abuse of notation) still ${V}^7$.
Eqns (\ref{3+1eqmot}) and (\ref{projconst}) imply that 
$ 
\dot{\bSigma}=|\bp|\,\big(\hbp\,\dot{t}-\dot{\br}\big).
$ 
Requiring $\bSigma=0$ is therefore consistent with the dynamics~:
the  motions of the chiral system lie in the intersection of the  $3$-dimensional characteristic leaves  of $V^9$ with the surface $V^7$ defined by spin enslaving; they remain therefore motions also for the extended dynamics  \footnote{ 
Alternatively, the restriction of the free two-form (\ref{freespinsigma})
 of $V^9$ to ${V}^7$ is (\ref{freechirsymp}).}. A chiral motion is embedded into $V^9$ by respecting the gauge condition (\ref{enslavement}), namely as ${\gamma}(t) = \big(\br(t)=\bx(t),\bp(t), \bs(t)=\half\,\hbp(t)\big)$ . 
%%%%%%%%%
\begin{figure}[h]
\begin{center}\vskip-3mm
\includegraphics[scale=.44]{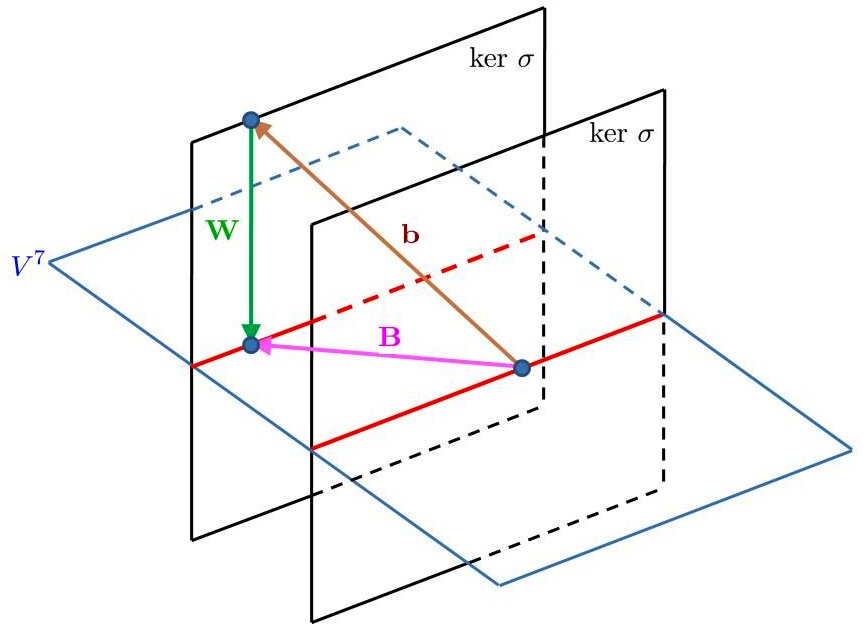}\vspace{-11mm}
\end{center}
\caption{\it Enslaving the spin amounts to embedding the chiral evolution space ${V^7}$
 into that, ${V^9}$,  of the massless spinning particle. A chiral motion 
 thus becomes the intersection of ${V}^7$
with the characteristic leaf of\ $\ker\sigma$ (depicted as a vertical plane).
Enslavement is not invariant under a (natural) boost $\bm{b}$, but a suitable
compensating Wigner-Souriau translation $\bm{W}$ allows us to
re-enslave the spin. The combination of the two transformations yields the twisted boost  $\bm{B}$ in (\ref{twistedboost}).
} 
\label{boostembedding}
\end{figure} 

%%%%%%%%%%%%%%%%%%%%%%%%%%%%%%%%%%%%%%%%
\section{Lorentz boost actions}\label{boostaction}

%%%%%%%%%%%%%%%%%%%%%%%%%%%%%%%%%%%%%%%%
%\subsection{Lorentz boosts acting on enslaved spin}
%%%%%%%%%%%%%%%%%%%%%%%%%%%%%%%%%%%%%%%%

We consider now an infinitesimal Lorentz boost, $\bbeta$. For $\bs=\half\hbp$ we have
$   
\delta_{\bbeta}(\bp\times\bs)=\half\,\bbeta\times\bp,
$  
which does \emph{not} vanish in general~: 
\emph{spin and momentum do not remain parallel} even if they were so initially: \emph{embedding  the chiral system into the Souriau model through spin enslavement is not boost-invariant~:} 
 \emph{natural boost symmetry is broken} by spin enslavement.
However, as dictated by the analogy with gauge symmetries \cite{FoMa},
let us apply an infinitesimal WS-translation 
\beq 
\delta_{\bW}\br=\bW,\;
\delta_{\bW}t=\hbp\cdot\bW,\;
\delta_{\bW}\bp=0,\;
\delta_{\bW}\bs=\bp\times\bW
\quad\hbox{with}\quad
\bW=\frac{\bbeta\times\hbp}{2|\bp|}\,.
\label{ensW}
\eeq  
Then $\delta_{\bW}\bs=\frac{1}{2}\big(\bbeta-\hbp\,(\hbp\cdot\bbeta)\big)$, implying that the combined transformation
$ 
\delta=\delta_{\bbeta}+\delta_{\bW}
$ 
 \emph{does preserve 
spin enslavement}, $\delta(\bp\times\bs)=0$.  The transformation of $V^7$ generated by $\delta$ 
is precisely the twisted boost (\ref{twistedboost})
\footnote{The order is irrelevant, since WS-shifts and boosts commute.}.

So far we studied  infinitesimal actions only.  But  our strategy is valid  also for finite transformations, as we show it now. Firstly, from Appendix C of \cite{DHchiral} we infer the action  of a finite boost  $\bb$ on the evolution space $V^9$, namely
\beq
\left\{
\barraynb{lll}
\br' &=& \br + (\gamma-1)\, (\hbb\cdot\,\br)\,\hbb + \gamma\,{t}\,\bb\,,
\\
t' &=& \gamma(\bb\cdot \br+ t)\, ,
\\
\bp' &=& \bp  + \gamma\,|\bp|\bb+ (\gamma-1)\, (\hbb\cdot\,\bp)\,\hbb\,,
\\
\bs'&=&
\bs 
+ (\gamma-1)\, \hbb\times( \bs \times \hbb)
 -
\gamma\, \bb \times\bSigma,
\earraynb\right.
\label{V9boost}
\eeq
which is consistent with the infinitesimal  action 
(\ref{infPVrtps}).
Then, starting with enslaved spin, $\bs=\half\hbp$, we find,  
\beq
\bSigma'
=
\frac{\bp'\times\bs'}{|\bp'|}
=
\half\big(\gamma^2|\bb|+(\gamma^2-1)\,\hbb\cdot\hbp\,\big)\,
\frac{\hbb\times\bp}{|\bp'|}
\label{finunchain}
\eeq
which does not vanish in general: spin is unchained. 
At last,  the finite WS-translation (\ref{finiteWigner}) with 
$ 
\bW ={\bSigma'}/{|\bp'|}
$ 
restores the validity of  (\ref{enslavement}): spin is  re-enslaved, $\bs''=\half\,{\hbp}'$.
The combined transformation for finite boosts,
\beq
\barraynb{lll}
\br'' &=& \br + \gamma\,{t}\,\bb + (\gamma-1)\, (\hbb\cdot\,\br)\,\hbb+\displaystyle\frac{\bSigma'}{|\bp'|},
\earraynb
\label{finitetwistedboost}
\eeq
completed with $t'' =t'$ and $\bp''=\bp'$ 
where $t',\,\bp'$ and $\bSigma'$ are given in (\ref{V9boost}) and (\ref{finunchain}), respectively.
The infinitesimal action is   (\ref{twistedboost}) 
as it should be.
In conclusion, Fig. \ref{boostembedding} is \emph{valid also for finite boosts}.

%%%%%%%%%%%%%%%%%%%%%%%%%%%%%%%%%%%%%%%%
%\subsection{Lorentz boosts acting on the space of motions}
%%%%%%%%%%%%%%%%%%%%%%%%%%%%%%%%%%%%%%%%

We now turn to the \emph{space of motions} \cite{SSD,DHchiral}, defined as
the quotient of the evolution space by the characteristic foliation of $\sigma$; we denote it by $M$. 
The equations of motion of the spin-extended system, (\ref{3+1eqmot}), imply that 
\beq
\tbx(t)=
\br(t)-\hbp\,{t}+\frac{\bSigma}{|\bp|}
\label{tx}
\eeq
is in fact a constant of the motion, $d\tbx/dt=0$. It
can be  used therefore to label the motion, i.e., a characteristic leaf in $V^9$. 
 The conserved momentum, $\bp$, is another good coordinate~set on $M$, which is  $6$-dimensional, and whose points can therefore be labeled by $\tbx$ and $\bp\neq0$. 

In \cite{DHchiral}, we derived the twisted boost (\ref{twistedboost}) from the Poincar\'e action on  the space of motions. Conversely, the latter can be obtained from our construction here.
 Boosting in $V^9$ according to (\ref{infPVrtps}) we find
$$
\delta_{\bbeta} \big(\br-\hbp{}t\big)=
-\hbp\,\big(\bbeta\cdot(\br-\hbp{}t)\big)
\quad\hbox{and}\quad
\displaystyle
\delta_{\bbeta}\left(\frac{\bSigma}{|\bp|}\right)
=
\frac{\bbeta\times\bs}{|\bp|}-\bSigma\,\frac{\bbeta\cdot\hbp}{|\bp|}\,,
$$
respectively. 
In view of (\ref{tx}), the preceding terms combine to yield 
\beq
\delta_{\bbeta}\tbx
=
\half\,\bbeta\times\frac{\hbp}{|\bp|}-\hbp\,(\bbeta\cdot{\tbx}).
\label{trajboost}
\eeq
Now WS-translations act in turn trivially on the space of motions, $\delta_{\bW}\tbx=\delta_{\bW}\bp=0$, since they move each characteristic leave within itself. 
Completing (\ref{trajboost}) with 
$ 
\delta_{\bbeta}\bp= |\bp|\bbeta
$ 
cf. (\ref{infPVrtps}), we end up with the (boost-)action on the space of motions, -- Eq. (3.19) in~\cite{DHchiral}. 
At last, applying (\ref{V9boost}) to each term in (\ref{tx}) allows us to confirm the finite action on the space of motions, Eq. \# (3.24) in \cite{DHchiral}.

%%%%%%%%%%%%%%%%%%%%%%%%%%%%%%%%%%%%%%%%
\section{The geometry of Wigner-Souriau transformations}\label{Wigner}
%%%%%%%%%%%%%%%%%%%%%%%%%%%%%%%%%%%%%%%%

Do WS-translations belong to the Poincar\'e group~? To answer this question, let us first briefly summarize Souriau's construction for ``elementary systems"
[meaning that the group acts symplectically and transitively] \cite{SSD}.

The connected component of the Poincar\'e group, $G$, 
can be identified with the group of matrices
$ 
g=\left(
\begin{array}{cc}
L&C\\
0&1
\end{array}\right)
$ 
where $L$ is a Lorentz transformation, $L\in{}H$ (the connected Lorentz group), and $C$ a Minkowski-space vector. 
The Poincar\'e Lie algebra, $\fg$, is hence spanned by the matrices
$ 
X=\left(
\begin{array}{cc}
\Lambda&\Gamma\\
0&0
\end{array}\right)
$ 
where the infinitesimal Lorentz transformations and translations, $\Lambda\in\so(3,1)$, and $\Gamma\in\bR^{3,1}$, respectively, are such that, in a given Lorentz frame,
$ 
\Lambda=\left(
\begin{array}{cc}
j(\bomega)&\bbeta\\
\bbeta^T&0
\end{array}\right)
$
and
$
\Gamma=\left(\begin{array}{c}\bgamma\\ \varepsilon\end{array}\right)
$ 
with $\bomega,\bbeta,\bgamma\in\bR^3$ interpreted as an infinitesimal rotation, boost, and space-translation, while $\varepsilon\in\bR$ stands for an infinitesimal time-translation. Here $j(\bomega)$ is the matrix of cross product in $3$-space, $j(\bomega)\bx=\bomega\times\bx$.

The Poincar\'e group acts on itself \emph{on the left},
$g \to hg$, for $h\in{}G$, 
and the quotient of $G$ by the subgroup of Lorentz transformations, $G/H$, can be identified with Minkowski space-time. This left-action of $G$ on itself projects
to the natural action, infinitesimally the two upper lines in (\ref{infPVrtps}), of the Poincar\'e group
on space-time.

The Poincar\'e group acts on the dual Lie algebra  by the co-adjoint representation, defined by $\Coad_g\mu\cdot{}X=\mu\cdot{}g^{-1}Xg$, for all $X\in\fg$. We denote here by $\mu=(M,P)\in\fg^*$ a ``moment'' of the Poincar\'e group, and by $\mu\cdot{}X=\half\,M_{\mu\nu}\Lambda^{\mu\nu}-P_\mu{}\Gamma^\mu$ its contraction with $X\in\fg$, see \cite{SSD}.

Contracting the Maurer-Cartan $1$-form, $g^{-1}dg$,
with and arbitrary fixed element $\mu_0$ of the dual of Lie algebra, yields a real one-form on the group $G$; we denote by 
$ %\beq
\sigma=d\big(\mu_0\cdot{}g^{-1}dg\big)
%\label{G2form}
$ %\eeq
its exterior derivative. As a general fact, the characteristic leaves of the two-form $\sigma$, defined by its kernel, are identified with the classical motions \cite{Duv} associated with $\mu_0$.
Indeed, the space of all motions, $M$, of an elementary system for the group $G$ is interpreted by Souriau~\cite{SSD} as the orbit of  some basepoint $\mu_0$ under the coadjoint action,
$ 
M=\Coad_G\mu_0\approx G/G_0,
$ 
where $G_0$ is the stability subgroup of 
$\mu_0$. The coadjoint orbit, $M$, is hence canonically endowed with the symplectic structure defined by the Kostant-Kirillov-Souriau two-form $\omega$, the image of $\sigma$ under the projection $G\to{}M$.

In our case and following \cite{SSD},
the basepoint can readily be chosen as $\mu_0=(M_0,P_0)$ with 
$ 
M_0=s\left(
\begin{array}{cc}
j(\hbp_0)&0\\
0&0
\end{array}\right)
$
and
$
P_0=|\bp_0|\left(\begin{array}{c}\hbp_0\\1\end{array}\right)
$ with $\hbp_0=\be_3$; here $s=\half$ is the scalar spin and the positive constant $|\bp_0|$ is the  energy.
 Then a straightforward calculation shows that the Poincar\'e Lie algebra element  $(\Lambda,\Gamma)$ with
 parameters $\bomega,\bbeta,\bgamma\in\IR^3,\varepsilon\in\IR$ leaves the chosen basepoint $\mu_0$ invariant whenever 
\begin{equation}
\bomega=\bbeta\times\hbp_0+\lambda\hbp_0,
\qquad
\bgamma=-\half\frac{\bbeta\times\hbp_0}{|\bp_0|}+\varepsilon\,\hbp_0.
\label{stab}
\end{equation} 
 The stability Lie algebra, $\fg_0$,
 is therefore $4$-dimensional, parametrized by  $(\bbeta,\varepsilon,\lambda)$, where $\bbeta\perp\hbp_0$, $\varepsilon\in\IR$, and $\lambda\in\IR$ represents an infinitesimal rotation around $\hbp_0$. We note that the evolution space $V^9$ is in fact the quotient
 of the Poincar\'e group by rotations around  $\hbp_0$,
 and~(\ref{infPVrtps}) above is in fact the projection 
 of the infinitesimal left-action of the group $G$ to $V^9\approx{}G/\SO(2)$, whereas the $2$-form %(\ref{G2form}) 
 $d\big(\mu_0\cdot{}g^{-1}dg\big)$ projects as (\ref{freespinsigma}), as
anticipated by the notation.
 
Remember now that the Poincar\'e group also acts on itself
\emph{from the right}, $g\to{}gh^{-1}$, for $h\in G$; its infinitesimal right-action is therefore given by matrix multiplication,
$ 
\delta_Xg=-gX,
$ 
where $X=\delta{h}$ at $h=1$.
Choosing in particular $X\in\fg_0$, the action on the group reads 
$ %\begin{equation}
\delta_{X}\left(
\begin{array}{cc}
L&C\\
0&1
\end{array}\right)
=
\left(
\begin{array}{cc}
-L\Lambda&-L\Gamma\\
0&0
\end{array}\right)
$ and
$
\Gamma=\left(\begin{array}{c}\bgamma\\
\varepsilon\end{array}\right),
%\label{deltaX0LambdaGamma}
$ %\end{equation}
where $\bgamma$ and $\bomega$ are as in (\ref{stab}).
This $4$-parameter vector field
 belongs, at each point $g\in G$, to the kernel of the two-form $\sigma$; in fact, it \emph{generates} its kernel \cite{SSD}. Renaming the Poincar\'e translation $C$ as $R=(\br,t)$, a space-time event, shows that the right-action on $G$ of a vector $X$ from the stability algebra $\fg_0$ yields, 
\begin{equation}
\delta_X\br
=
\half\,\frac{\bbeta\times\hbp}{\vert\bp\vert}-\varepsilon\,\hbp
\qquad
\&
\qquad
\delta_X t=-\varepsilon.
\label{deltabrt}
\end{equation}
This transformation satisfies the condition $\hbp\cdot \delta_X\br=\delta_Xt$ required for an infinitesimal WS-translation;  
conversely, any WS-translation $\bW$ is of the form (\ref{deltabrt}), with  $\bbeta$ perpendicular to~$\bp$
and $\lambda$ arbitrary. Therefore (with a slight abuse) we will refer to $\fg_0$ acting from the right as a WS-translations.

Comparing now (\ref{deltabrt}) with (\ref{ensW}) allows us to conclude that 
\emph{while a boost $\bbeta$ acting on the left unchains the spin, the latter is re-enslaved by a WS-translation with the same boost, $\bbeta$, acting from the right}.

It is readily verified that the right-action of an $X$ from the stability algebra $\fg_0$ acts as $\delta_X\bp=0$ and 
$\delta_X\bs=\half\,\hbp\times(\hbp\times\bbeta)$, consistently with the infinitesimal action.  
Thus, after rotations around~$\hbp_0$ are factored out, the right-action of the stability subalgebra $\fg_0$
projects to $V^9$ as the WS trans\-lations. 
In conclusion, Wigner-Souriau translations \emph{are}
Poincar\'e transformations, --- but which act on the 
\emph{right} and not on the left as natural ones do.
Projecting further down to the space of motions, $M$, they act trivially. 
The various spaces are shown, symbolically, on Fig. \ref{diagram} below.
%%%%%%%%%
\begin{figure}[h]
\begin{center}%\vskip-6mm
\includegraphics[scale=.36]{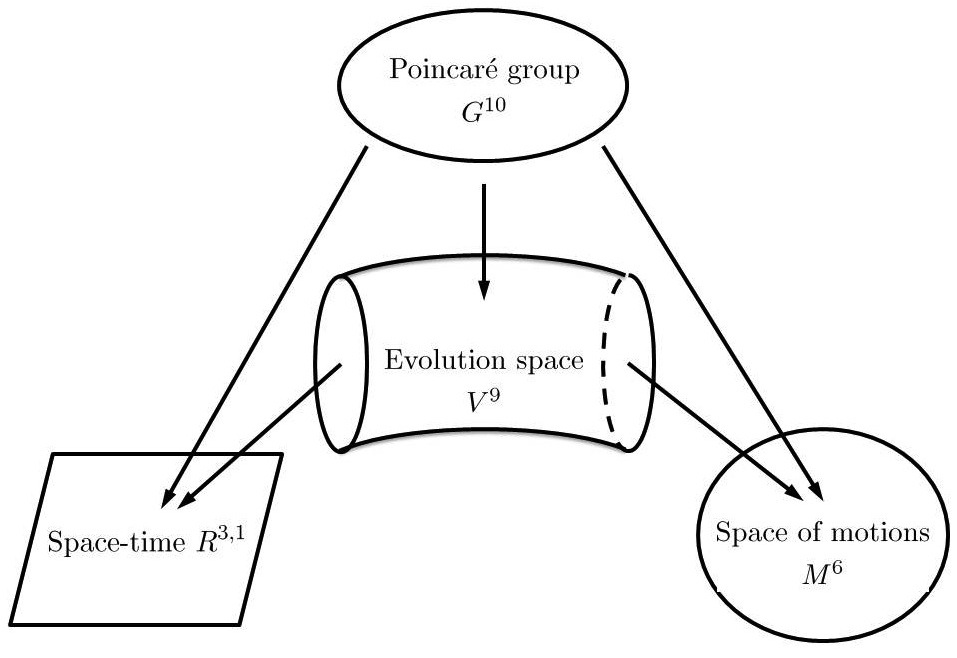}
\vspace{-10mm}
\end{center}
\caption{\it The evolution space, $V^9$, is the quotient of
the Poincar\'e group, $G^{10}$ by rotations around $\hbp_0$. The space of motions is identified with a coadjoint orbit 
$M^6\approx G^{10}/G_0$ of $G^{10}$, where $G_0$, the stability group of the basepoint $\mu_0$. The left-action of $G^{10}$ on itself projects to the natural Poincar\'e action on both the space of motions and Minkowski space-time, consistently with projecting first
to~$V^9$ and then to $M^6$ and $\IR^{3,1}$, respectively. 
}
\label{diagram}
\end{figure}
 
%%%%%%%%%%%%%%%%%%%%%%%%%%%%%%%%%%%%%%
\section{Relation to non-commutativity}
%%%%%%%%%%%%%%%%%%%%%%%%%%%%%%%%%%%%%%

The spaces of motions of both the chiral and the Souriau systems carry the symplectic structure $\omega$ in
(\ref{freechirsymp}) [with $\tbx$ replacing $\bx$] \cite{DHchiral}. Translating into Poisson bracket language, we have
$ 
\{\xi^\alpha,\xi^\beta\}=\omega^{\alpha\beta}(\xi)
$ 
where $(\omega^{\alpha\beta})=-(\omega_{\alpha\beta})^{-1}$ are the coefficients of the inverse of the  symplectic form 
$\omega=\half\omega_{\alpha\beta}(\xi)d\xi^\alpha\wedge{}d\xi^\beta$. Therefore the components of the space of motions coordinates $\tbx$ do \emph{not to Poisson-commute}, 
%%%%
\beq
\big\{\tilde{x}^i,\tilde{x}^j\big\}=\frac12\,\epsilon^{ijk}\frac{p_k}{|\bp|^3}\,.
\label{NC}
\eeq
Putting $t=0$ and $\bSigma=0$ into (\ref{tx}) shows that the chiral coordinate $\bx$ satisfies the same commutation relation.
Non-commutativity is thus a consequence of spin, cf. \cite{Barut,Bacry,DE,SpinOptics}; 
chiral fermions provide just another example of non-commutative mechanics.

It is worth stressing that the non-commutativity is \emph{unrelated} to WS translations. Non-commutativity arises in fact due to the twisted symplectic structure of the \emph{space of motions} (modelled by the phase space); but the WS translations act on the evolution space \emph{before projecting, trivially}, to the space of motions.

In the operator context a \emph{Bopp-shift} transforms the commutation relations to canonical ones \cite{Bopp}. 
 In the (semi)classical context we are interested in it amounts to introducing new, canonical coordinates, eliminating spatial non-commutativity altogether. Now Darboux's theorem \cite{Darboux}
guarantees that this is always possible \emph{locally}.
In detail, we redefine the coordinates as
\begin{equation}
\tbX=\tbx+\ba(\bp),
\qquad
\tbP=\bp,
\label{Darbouxgeneral}
\end{equation}
where $\ba (\bp)$ is a ``Berry'' vector potential. Then,  in terms of $\tbP$ and $\tbX$, the symplectic form~$\omega$ in (\ref{freechirsymp})  has the
canonical expression $\omega=d\widetilde{P}_i\wedge{}d\widetilde{X}^i$, so that  
$\{\widetilde{X}^i,\widetilde{X}^j\}=
\{\widetilde{P}_i,\widetilde{P}_j\}=0$ and 
$\{\widetilde{P}_i,\widetilde{X}^j\}=\delta_i^j$.
 Spin seems thus eliminated.
This is \emph{not} possible globally, though, since
 any~$\ba$ is singular
along a Dirac string, as it is well-known.

The variation of the Berry vector potential under a boost $\bbeta\in\IR^3$ is then found to be
$ 
\delta\ba=-\bbeta\times \displaystyle\frac{\hat{\bp}}{2|\bp|} -(\bm{a}\cdot\bbeta)\,\hat{\bp}
+
\bnabla_{\bp}\big(|\bp|\,\bbeta\cdot\ba\big).
$ 
It follows that, in terms of the Darboux coordinates (\ref{Darbouxgeneral}),  the infinitesimal action of boosts is given by
\beqa
\delta\tbX=
-(\bbeta\cdot\widetilde{\bm{X}})\,\frac{\tbP}{|\tbP|} +
\bnabla_{\tbP}\big(|\tbP|\,\bbeta\cdot\ba\big),
\qquad
\delta\tbP=\bbeta|\tbP|.
\label{BoostsDarboux}
\eeqa
Redefining analogously the chiral coordinate as
$
\bX=\bx+\bm{a}
$
and $\bP=\bp$
 allows us to infer
\begin{eqnarray}
\label{Xboost1}
\delta\bX=\bbeta t-(\bbeta\cdot\bm{a})\hat{\bP}+\bnabla_\bP(|\bP|\,\bbeta\cdot\ba)=
\bbeta t+
|\bP|\bnabla_\bP\big(\bbeta\cdot\ba\big),
\qquad
\delta\bP=\bbeta|\bP|,
\end{eqnarray}
which is reminiscent of but still different from the natural action (\ref{infPVrtps}).

%%%%%%%%%%%%%%%%%%%%
\section{Conclusion}
%%%%%%%%%%%%%%%%%%%% 

In this Letter, we re-derived 
the twisted Lorentz symmetry (\ref{twistedboost}) of chiral fermions by embedding the theory of Refs. \cite{Stephanov,SonYama2,Stone,Stone3,ChenSon}
into Souriau's massless spinning model \cite{SSD,DHchiral} by spin enslavement, (\ref{enslavement}), viewed as a gauge fixing. The latter is not boost invariant,
but enslavement can be restored by a suitable compensating WS-translation, which is analogous to  a gauge transformation \cite{FoMa}. Our formula 
(\ref{finitetwistedboost})
extends  (\ref{twistedboost}) to finite transformations.

The motions of the extended model are not mere curves but 3-planes, swept by WS-translations. The massless relativistic particle is delocalized  \cite{Wigner,Trautman,Penrose,SSD,Skagerstam,Stone3,DHchiral} and behaves rather as a \emph{3-brane}. 

Coupling  to an external electromagnetic field breaks the  WS ``gauge'' symmetry, so that spin can not be enslaved. The motions then take place along curves: the particle gets localized  \cite{DHchiral}.
New results \cite{Stone4} indicate however
that while a WS translation \emph{is unobservable} for an
infinite plane wave, it \emph{is observable} for a wave packet. Intuitively, forming and keeping together a wave packet behaves as a sort of interaction.

\begin{acknowledgments} 
We would like to thank Mike Stone for calling our attention at chiral fermions and for enlightening correspondence.
 ME is indebted to \textit{Service de Cooperation et Culturel de l'Ambassade de France en Turquie} and to the \emph{Laboratoire de Math\'ematiques et de Physique Th\'eorique} of Tours University.  P. Kosinski and his collaborators work on a related project \cite{Kosinski}.
 This work was supported by the Major State Basic Research Development Program in China (No.
2015CB856903) and the National Natural Science Foundation of
China (Grant No. 11035006 and 11175215).
\end{acknowledgments}
%%%%

\goodbreak

%%%%%%%%%%%%%%%%%%%%%%%%%%%%%%%%%%%%%%%%%%%%%%%%%%%%%%%%%%%%%%%%%%%%%%%%%%%%%%
%%%%%%%%%%%%%%%%%%%%%%%%%%%%%%%%%%%%%%%%%%%%%%%%%%%%%%%%%%%%%%%%%%%%%%%%%%%%%%

\end{document}